\documentclass[twocolumn,superscriptaddress,amsmath,amssymb]{revtex4}
\usepackage{graphicx}% Include figure files
\usepackage{dcolumn}% Align table columns on decimal point
\usepackage{bm}% bold math
\usepackage{color}

\usepackage[colorlinks,
           bookmarks=true,
           linkcolor=blue,
           urlcolor=blue,
            anchorcolor=black,
            citecolor=blue
            ]{hyperref}

\usepackage[all]{hypcap}

\begin{document}

\title{Correlation functions of net-proton multiplicity distributions in Au + Au collisions at energies available at the BNL Relativistic Heavy Ion Collider from a multiphase transport model}

\author{Yufu Lin }
\affiliation{Key Laboratory of Quark and Lepton Physics (MOE) and Institute of Particle Physics, \\
Central China Normal University, Wuhan 430079, China}
\author{Lizhu Chen}
\affiliation{School of Physics and Optoelectronic Engineering, \\
Nanjing University of Information Science and Technology, Nanjing 210044, China}
\author{Zhiming Li}
\email{lizm@mail.ccnu.edu.cn}
\affiliation{Key Laboratory of Quark and Lepton Physics (MOE) and Institute of Particle Physics, \\
Central China Normal University, Wuhan 430079, China}

\begin{abstract}
Fluctuations of conserved quantities are believed to be sensitive observables to probe the signature of the QCD phase transition and critical point. It was argued recently that measuring the genuine correlation functions (CFs) could provide cleaner information on possible nontrivial dynamics in heavy-ion collisions.With the AMPT (a multiphase transport) model, the centrality and energy dependence of various orders of CFs of net protons in Au + Au collisions at $\sqrt{s_\mathrm{NN}}$=7.7, 11.5, 19.6, 27, 39, 62.4 and 200 GeV are investigated. The model results show that the number of antiprotons is important and should be taken into account in the calculation of CFs at high energy and/or in peripheral collisions. It is also found that the contribution of antiprotons is more important
for higher order correlations than for lower ones. The CFs of antiprotons and mixed correlations play roles comparable to those of protons at high energies. Finally, we make comparisons between the model calculation and experimental data measured in the STAR experiment at the BNL Relativistic Heavy Ion Collider.
\end{abstract}

\maketitle

\section{Introduction}

The search for the structure of the QCD phase diagram is one of the main goals in relativistic heavy-ion collisions~\cite{StephanovPD,adams2005experimental}. Event-by-event fluctuations of conserved quantities are less affected by final state interactions in the hadronic phase
and thus have been believed to be sensitive to the QCD critical point~\cite{conservecharge0,conservecharge1,ejiri2006hadronic,conservecharge2,conservecharge3,conservecharge4,asakawa2009third}. With the data collected from the first beam
energy scan (BES-I) program at the BNL Relativistic Heavy
Ion Collider (RHIC), cumulants of net-proton, net-charge,
and net-kaon multiplicity distributions have been measured~\cite{net_proton2010,net_proton2014,net_charge2014,xu2017_net-kaon,amal2017_net-kaon,STARnet_kaon2017,Jochen_star,phenix}. A nonmonotonic energy dependence of the cumulant ratio of the fourth order to the second order ($\kappa \sigma^{2}$) for net protons in central Au + Au collisions has been observed in the
STAR experiment, with a minimum value around 19.6 GeV and strong enhancements at 7.7 and 11.5 GeV~\cite{luo2015energy,xiaofenglongpaper}.

Recently, it was argued~\cite{BLing,ab1,ab2,ab22,ab3} that measuring multiparticle
correlation function could provide cleaner information on the dynamics of critical fluctuations in heavy-ion collisions, since cumulants may mix correlations of different orders. By ignoring the contributions from antiprotons, various orders of correlation functions between protons can be extracted from the cumulants measured by the STAR experiment. It was found that the strong enhancement of fourth-order cumulant at 7.7 GeV is due to large four-particle correlations, while the negative two-particle correlations are dominant at 19.6 GeV~\cite{ab2}. With removal of the four-particle correlation function from the cumulants, i.e., only the two- and three-particle correlation functions remain, the nonmonotonic behavior
observed in the fourth-order proton cumulant disappears~\cite{Roli}. The ultrarelativistic quantum molecular dynamics (UrQMD) model cannot explain the observed behaviors of the proton correlation functions from STAR~\cite{luorqmd}. The correlation functions could also be effectively used to study the long-range correlations of the system~\cite{bzdak2016multiparticle} and describe the asymmetric component of rapidity correlations measured by the ATLAS experiment~\cite{asymmetry1,asymmetry2}.

It should be pointed out that proton number is not a conserved charge although the number of antiprotons is comparably small and may be neglected at very low RHIC
energies. With increasing collision energies, the experimentally measured antiproton to proton ratio ($\bar{p}/p$) increases dramatically ~\cite{pbarpratio}, and we should consider the contribution from antiprotons as well as that from protons when calculating cumulants or correlation functions. On the other hand, the mixed correlation functions between protons and antiprotons may carry important information about the system under investigation. In Ref.~\cite{mix} it is suggested that measurement of these mixed correlations could help to identify the possible origination of proton clustering, which can be used to qualitatively understand the preliminary STAR results for multiparticle correlation functions of different orders.

In this work, we plan to perform a study on the correlation functions of net protons by taking both protons and antiprotons into account. The correlation functions of net protons at various
RHIC BES energies and centralities are investigated and compared to those of only protons. We also make comparisons of correlation functions from protons or antiprotons to those from
mixed correlations. The contributions of antiproton numbers to various orders of correlation functions are systematically studied.

The paper is organized as follows. In Sec. II, we introduce the observables and relations between cumulant, factorial moment, and correlation function. A brief introduction of the
AMPT (a multiphase transport) model is given in Sec. III. Then, the centrality dependence of various orders of correlation functions of net protons, only protons or antiprotons, and mixed correlation are investigated by the AMPT model in Au + Au collisions at $\sqrt{s_\mathrm{NN}}$=7.7 to 200 GeV. Finally, we give a summary of this work.

%%%%%%%%%%%%%%%%%%%%%%%%%%%%%%%%%%%%%%%%%%%%%%%%%%%%%%%%%%%%%%%%%%%%%%%%%%%%%%%
\section{Cumulant, Factorial Moment and Correlation Function}

\begin{figure*}[!htb]
\hspace{-0.8cm}
\includegraphics[scale=0.92]{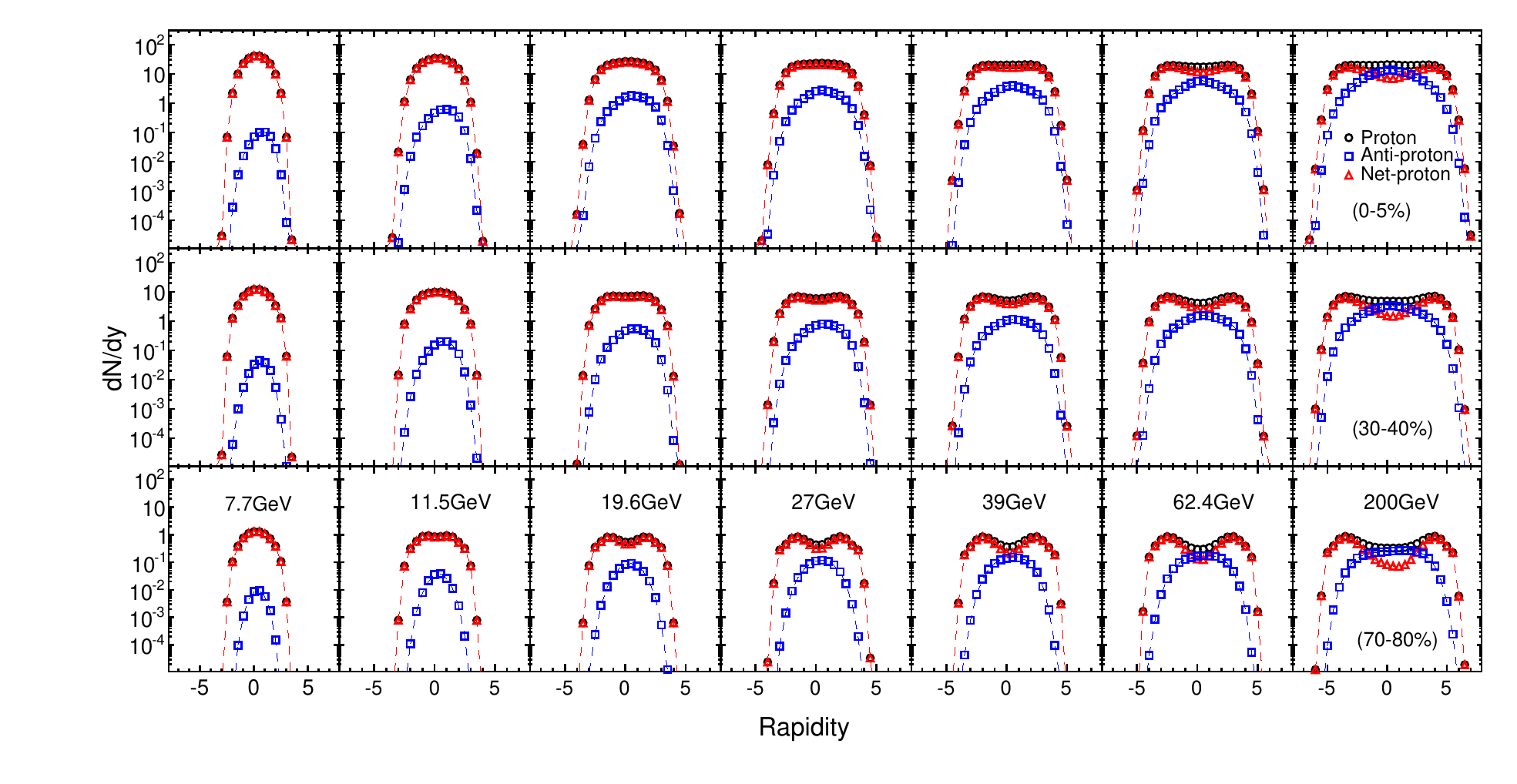}
\caption{The rapidity distributions ($dN/dy$) of protons, anti-proton and net-proton multiplicities in Au+Au collisions at $\sqrt{s_{NN}}$ = 7.7, 11.5, 19.6, 27, 39, 62.4 and 200 GeV for three centrality bins (0-5\%, 30-40\%, 70-80\%).}
\label{Fig:dndy}
\end{figure*}

In statistics, the distribution function of a measurement can be characterized by various orders of cumulants. The $n$th order cumulants ($C_{n}$) of a multiplicity distribution can be defined as:\\
\parbox{7.cm}{
\begin{eqnarray*}
     C_{1} &=& \langle N\rangle \\ %N_{p} - N_{\bar{p}}
     C_{2} &=& \langle (\delta N)^2 \rangle\\
     C_{3} &=& \langle (\delta N)^3 \rangle \\
     C_{4} &=& \langle (\delta N)^4 \rangle-3 \langle (\delta N)^2 \rangle^2
\end{eqnarray*}}\hfill
\parbox{1cm}{ \begin{equation}\label{cumulant} \end{equation}}\\
where $N$ represents the number of multiplicities measured in one event and $\delta N= N- \langle N\rangle$ denotes the deviation of $N$ to its mean value. The measurement $N$ could be the number of protons, anti-protons or net-protons.

In order to cancel the volume effect, the ratio of the fourth order to the second order cumulants, $\kappa \sigma^{2}$, is often used to study the non-monotonic behavior in current heavy-ion experiments~\cite{net_proton2010,net_proton2014,net_charge2014}. It can be obtained by:\\
\begin{equation}
%\begin{align*}
	\kappa\sigma^2	=	\frac{C_4}{C_2}
 \label{Eq:KV}
%\end{align*}
\end{equation}

Suppose that the final state particles are characterized by the multiplicity distribution $P(N)$, where $N$ is only one species of particles, such as protons. The $n$-th order factorial moment $F_{n}=\langle N!/(N-n)!\rangle$ can be calculated from the generating function $H(z)$:\\
\begin{equation}
	F_{n}=\frac{d^{n}}{dz^{n}}H(z)|_{z=1}, H(z)=\sum_{N}P(N)z^{N}.
 \label{Eq:FH}
\end{equation}

The $n$-th order correlation function, $\hat{\kappa}_{n}$, is then given by analogous derivatives of the logarithm of $H(z)$,\\
\begin{equation}
	\hat{\kappa}_{n}=\frac{d^{n}}{dz^{n}}\ln[H(z)]|_{z=1}.
 \label{Eq:KH}
\end{equation}

The correlations functions can be obtained from the factorial moments:\\
%\parbox{6.5cm}{
\begin{eqnarray}
    \hat{\kappa}_{1}&=&F_{1} \nonumber\\
    \hat{\kappa}_{2}&=&F_{2}-F_{1}^{2} \nonumber\\
    \hat{\kappa}_{3}&=&F_{3}-3F_{1}F_{2}+2F_{1}^{3}\\
    \hat{\kappa}_{4}&=&F_{4}-4F_{1}F_{3}-3F_{2}^{2}+12F_{1}^{2}F_{2}-6F_{1}^{4} \nonumber
\end{eqnarray}
%}\hfill
%\parbox{1cm}{ \begin{equation}\label{cumulant2} \end{equation}}.\\

Cumulants can also be expressed in terms of the factorial moments~\cite{ab1,ab3}. Thus, cumulants and correlation functions have direct connections with each other:
\begin{eqnarray}\label{Eq:proton}
    C_{2} &=& \langle N\rangle+\hat{\kappa}_{2}  \label{Eq:p_C2}\\
    C_{3} &=& \langle N\rangle+3\hat{\kappa}_{2}+\hat{\kappa}_{3} \label{Eq:p_C3}\\
    C_{4} &=& \langle N\rangle+7\hat{\kappa}_{2}+6\hat{\kappa}_{3}+\hat{\kappa}_{4}  \label{Eq:p_C4}
\end{eqnarray}\\

It can be clearly seen that  cumulants mix correlations of different orders. For example, from Eq.~\eqref{Eq:p_C4} we find that the forth-order cumulant includes the second-, third- and fourth-order correlation functions.

We must point out that the above relationships are valid only when one type of particle, such as the proton, is taken into account. For net-protons, i.e. both protons and anti-protons are considered, the relationships between correlation functions and factorial moments can be deduced~\cite{ab1,relation1, relation2}:
%\begin{widetext}
%\parbox{6.5cm}{
\begin{align}
    \hat{\kappa}_{2}^{(2,0)}=&F_{2,0}-F_{1,0}^{2} \nonumber\\ % ,\;\;\;
    \hat{\kappa}_{2}^{(0,2)}=&F_{0,2}-F_{0,1}^{2} \nonumber\\ %   ,\;\;\;
    \hat{\kappa}_{2}^{(1,1)}=&F_{1,1}-F_{1,0}F_{0,1}  \nonumber\\ %  ,\;\;\;
    \hat{\kappa}_{3}^{(3,0)}=&F_{3,0}-3F_{1,0}F_{2,0}+2F_{1,0}^{3} \nonumber\\ %   ,\;\;\;
    \hat{\kappa}_{3}^{(0,3)}=&F_{0,3}-3F_{0,1}F_{0,2}+2F_{0,1}^{3}  \\ %  ,\;\;\;
    \hat{\kappa}_{3}^{(2,1)}=&F_{2,1}-F_{0,1}F_{2,0}-2F_{1,0}F_{1,1}+2F_{0,1}F_{1,0}^{2}\nonumber\\
    \hat{\kappa}_{3}^{(1,2)}=&F_{1,2}-F_{1,0}F_{0,2}-2F_{0,1}F_{1,1}+2F_{1,0}F_{0,1}^{2} \nonumber\\
    \hat{\kappa}_{4}^{(4,0)}=&F_{4,0}-4F_{1,0}F_{3,0}-3F_{2,0}^{2}+12F_{1,0}^{2}F_{2,0}-6F_{1,0}^{4}\nonumber\\
    \hat{\kappa}_{4}^{(0,4)}=&F_{0,4}-4F_{0,1}F_{0,3}-3F_{0,2}^{2}+12F_{0,1}^{2}F_{0,2}-6F_{0,1}^{4}\nonumber\\
    \hat{\kappa}_{4}^{(3,1)}=&F_{3,1}-F_{0,1}F_{3,0}-3F_{1,0}F_{2,1}-3F_{1,1}F_{2,0}\nonumber\\
            &+6F_{0,1}F_{1,0}F_{2,0}+6F_{1,0}^{2}F_{1,1}-6F_{0,1}F_{1,0}^{3}\nonumber\\
    \hat{\kappa}_{4}^{(1,3)}=&F_{1,3}-F_{1,0}F_{0,3}-3F_{0,1}F_{1,2}-3F_{1,1}F_{0,2}\nonumber\\
            &+6F_{1,0}F_{0,1}F_{0,2}+6F_{0,1}^{2}F_{1,1}-6F_{1,0}F_{0,1}^{3}\nonumber\\
    \hat{\kappa}_{4}^{(2,2)}=&F_{2,2}-2F_{0,1}F_{2,1}+(2F_{0,1}^{2}-F_{0,2})F_{2,0} -2F_{1,0}F_{1,2}\nonumber\\
            &-2F_{1,1}^{2}+8F_{0,1}F_{1,0}F_{1,1}+(2F_{0,2}-6F_{0,1}^{2})F_{1,0}^{2}\nonumber
\end{align}
%}\hfill
%\parbox{1cm}{ \begin{equation}\label{cumulant} \end{equation}}
%\end{widetext}
where $F_{i,k}\equiv \langle\frac{N!}{(N-i)!}\frac{\bar{N}!}{(\bar{N}-k)!}\rangle$, and $i$ is the number of protons and $k$ is that of anti-protons.  $\hat{\kappa}_{n}^{(i,k)}$ represents the $n$-th order correlation function with $i$ protons and $k$ anti-protons. We call it a mixed correlation function if both $i$ and $k$ are nonzero. Then one can deduce the formulas between cumulants and correlation functions for net-protons~\cite{relation1, relation2}:\\
\parbox{6.5cm}{
\begin{eqnarray}
% \nonumber to remove numbering (after each equation)
  C_{2} &=& \langle N\rangle+\langle\overline{N}\rangle+\hat{\kappa}_{2}^{(2,0)}+\hat{\kappa}_{2}^{(0,2)}-2\hat{\kappa}_{2}^{(1,1)}  \label{Eq:net_C2}\\
  C_{3} &=& \langle N\rangle-\langle\overline{N}\rangle+3\hat{\kappa}_{2}^{(2,0)}-3\hat{\kappa}_{2}^{(0,2)}+\hat{\kappa}_{3}^{(3,0)}-\hat{\kappa}_{3}^{(0,3)} \nonumber\\
  &&-3\hat{\kappa}_{3}^{(2,1)}+3\hat{\kappa}_{3}^{(1,2)}  \label{Eq:net_C3}\\
  C_{4} &=& \langle N\rangle+\langle\overline{N}\rangle+7\hat{\kappa}_{2}^{(2,0)}+7\hat{\kappa}_{2}^{(0,2)}-2\hat{\kappa}_{2}^{(1,1)}+6\hat{\kappa}_{3}^{(3,0)} \nonumber
  \\&&+6\hat{\kappa}_{3}^{(0,3)}-6\hat{\kappa}_{3}^{(2,1)}-6\hat{\kappa}_{3}^{(1,2)}+\hat{\kappa}_{4}^{(4,0)}+\hat{\kappa}_{4}^{(0,4)} \nonumber
  \\&&-4\hat{\kappa}_{4}^{(3,1)}-4\hat{\kappa}_{4}^{(1,3)}+6\hat{\kappa}_{4}^{(2,2)}  \label{Eq:net_C4}
\end{eqnarray}}\\

In order to simplify notations used in the discussions on the contributions of different order correlation functions to the fourth-order cumulant in Sec.IV, we make the following definitions according to Eq.~\eqref{Eq:p_C4} for protons and \eqref{Eq:net_C4} for net-protons, respectively:\\
\begin{align}
	&(\hat{\kappa}_{2})_{p} = 7\hat{\kappa}_{2} \nonumber\\ %   ,\;\;\;
    &(\hat{\kappa}_{3})_{p} = 6\hat{\kappa}_{3} \nonumber\\ %  ,\;\;\;
    &(\hat{\kappa}_{4})_{p} = \hat{\kappa}_{4}  \label{Eq:correlation}\\
    &(\hat{\kappa}_{2})_{net}  =  7\hat{\kappa}_{2}^{(2,0)}+7\hat{\kappa}_{2}^{(0,2)}-2\hat{\kappa}_{2}^{(1,1)}  \nonumber\\
    &(\hat{\kappa}_{3})_{net} =  6\hat{\kappa}_{3}^{(3,0)}+6\hat{\kappa}_{3}^{(0,3)}-6\hat{\kappa}_{3}^{(2,1)}-6\hat{\kappa}_{3}^{(1,2)} \nonumber \\
    &(\hat{\kappa}_{4})_{net} =   \hat{\kappa}_{4}^{(4,0)}+\hat{\kappa}_{4}^{(0,4)}-4\hat{\kappa}_{4}^{(3,1)}-4\hat{\kappa}_{4}^{(1,3)}+6\hat{\kappa}_{4}^{(2,2)} \nonumber
\end{align}
where $(\hat{\kappa}_{n})_{p}$  represents the $n$th-order correlation function for only protons and $(\hat{\kappa}_{n})_{net}$ is that for net-protons. The scaled numbers in the formulas of $(\hat{\kappa}_{n})_{p}$ and $(\hat{\kappa}_{n})_{net}$ are obtained from Eq.~\eqref{Eq:p_C4} for protons and \eqref{Eq:net_C4} for net-protons. These numbers reflect the relative contribution of different correlation functions to the fourth-order cumulant.

From the above definitions, we can infer that $(\hat{\kappa}_{n})_{net}$ will degrade to be equal to $(\hat{\kappa}_{n})_{p}$ if we omit the contributions of anti-protons. Therefore, in the real data sample, if the measured values of $(\hat{\kappa}_{n})_{p}$ and $(\hat{\kappa}_{n})_{net}$ have a clear difference, it means the contribution of anti-protons is important and should be taken into account when calculating correlation functions in that case.

\begin{figure*}[!htpb]
\hspace{-0.8cm}
\includegraphics[scale=0.92]{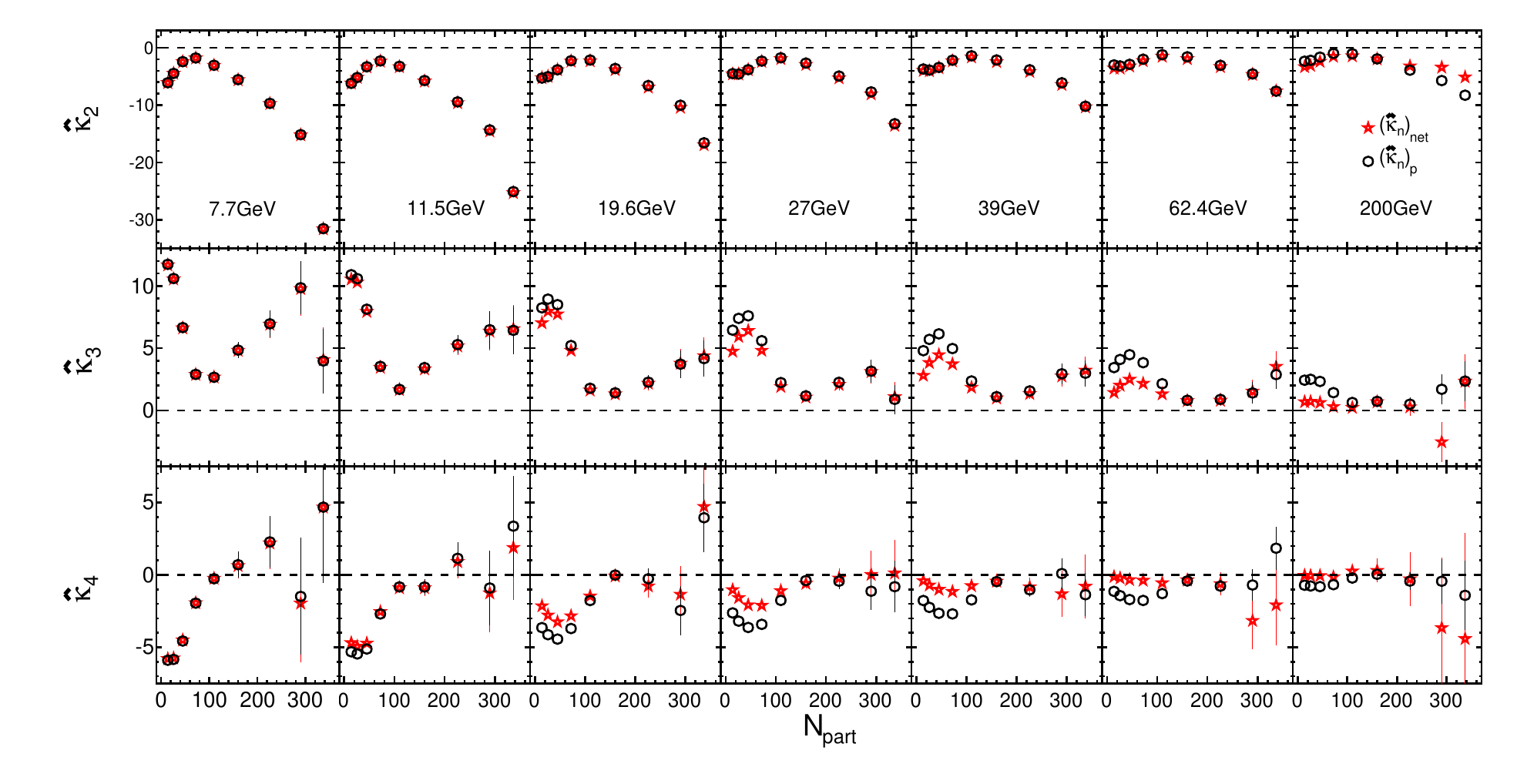}
\caption{Centrality dependence of the second-, third- and fourth-order correlation functions of net-protons and protons in Au+Au collisions at $\sqrt{s_{NN}}$ = 7.7, 11.5, 19.6, 27, 39, 62.4 and 200 GeV.}
\label{Fig:Net_CF}
\end{figure*}

%%%%%%%%%%%%%%%%%%%%%%%%%%%%%%%%%%%%%%%%%%%%%%%%%%%%%%%%%%%%%%%%%%%%%%%%%%%%%%%
\section{AMPT Model}
A multiphase transport model with string melting (AMPT-SM)is a transport model including four main processes: the initial condition, the partonic interactions, the conversion from partonic matter to hadronic matter, and the hadronic interactions. The initial condition is obtained from the HIJING model, which includes the spatial and momentum distributions of minijet partons and soft string excitation. Then all the excited strings convert to partons. Scattering among partons is modeled by Zhang¡¯s parton cascade (ZPC), which at present includes only two-body scattering with cross sections obtained from pQCD with screening masses. A simple quark coalescence model based on the quark spatial information is used to combine partons into hadrons. The dynamics of the subsequent hadronic matter is described by a relativistic transport (ART) model. More details about the AMPT model can be found in Refs~\cite{lin2005AMPT,lin2014recent}.

In this paper, we perform our calculations with AMPT-SM model in version 2.21 for Au+Au collisions at $\sqrt{s_{NN}}$=7.7, 11.5, 19.6, 27, 39, 62.4, 200 GeV and the corresponding statistics are($\times 10^6$) 23.7, 27.6, 22.5, 21.6, 18.5, 10.5, 8.8 , respectively.

%%%%%%%%%%%%%%%%%%%%%%%%%%%%%%%%%%%%%%%%%%%%%%%%%%%%%%%%%%%%%%%%%%%%%%%%%%%%%%%
\section{Results and Discussions}
In the AMPT model calculations, we apply the same kinematic cuts and technical analysis methods as those used for the STAR experiment data~\cite{luo2015energy}. The protons and anti-protons are measured at mid-rapidity ($|y|<0.5$) and within the transverse momentum range $0.4<p_{T}<2.0$ GeV/c. The centrality is defined by the charged pion and kaon multiplicities within pseudo-rapidity $|\eta|<1.0$, which can effectively avoid auto-correlation effects in the measurement of cumulants and correlation functions. In order to suppress the volume fluctuations originating from the finite centrality bin width, we apply the centrality bin width correction~\cite{CBWC} to the measurement. The statistics error is estimated by the bootstrap method~\cite{efron1993introductionBootstrap,efron1986introductionBootstrap}.

Figure \ref{Fig:dndy} displays the normalized rapidity distribution ($\mathrm{d}N/\mathrm{d}y$) of protons, anti-protons and net-protons in the most central (0-5\%), mid-central (30-40\%) and peripheral (70-80\%) Au+Au collisions at $\sqrt{s_\mathrm{NN}}$=7.7 to 200 GeV. With collision energy increase, the $dN/dy$ distributions of protons in the  central rapidity region monotonically decrease, while those of anti-proton increase with increasing energies.  If energy is fixed, the yield of anti-protons is found to be closer to that of protons in more peripheral collisions. Due to the negligible production of anti-protons in the  low energies, the $dN/dy$ distribution at central rapidity region of net-proton is observed to closely follow that of protons, while a clear difference between them can be found at high energies, especially at 200 GeV. These can be explained by the different particle production mechanisms at different RHIC energies. Baryon stopping is more important at low energies, while  pair production dominates the production of protons and anti-protons at high energies.

\begin{figure*}[!htpb]
\hspace{-0.8cm}
\includegraphics[scale=0.75]{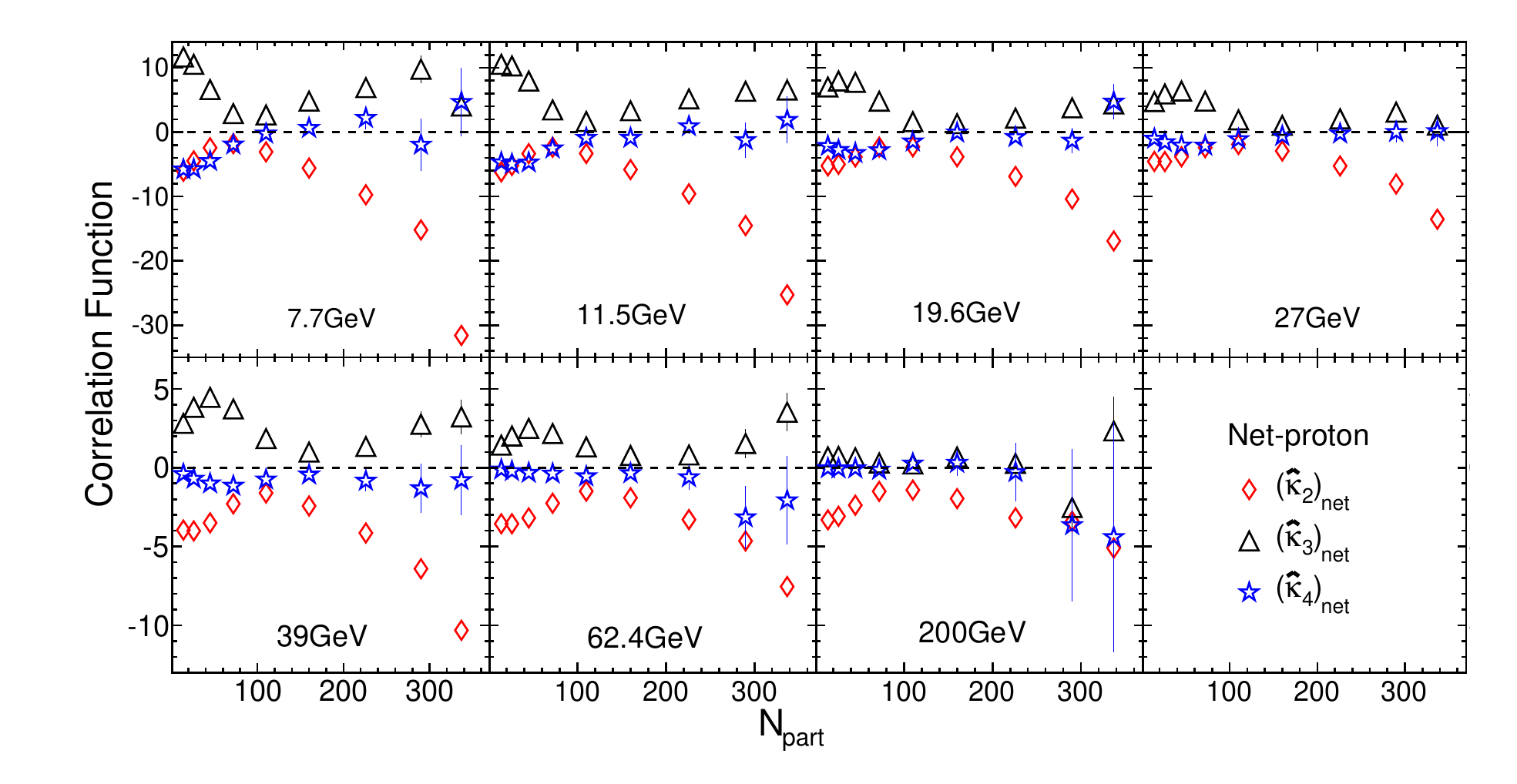}
\caption{Centrality dependence of various orders of correlation functions of net-protons in Au+Au collisions at $\sqrt{s_{NN}}$ = 7.7, 11.5, 19.6, 27, 39, 62.4 and 200 GeV.}
\label{Fig:p&&Net_CF}
\end{figure*}

\begin{figure*}[!htp]
\hspace{-0.8cm}
\includegraphics[scale=0.92]{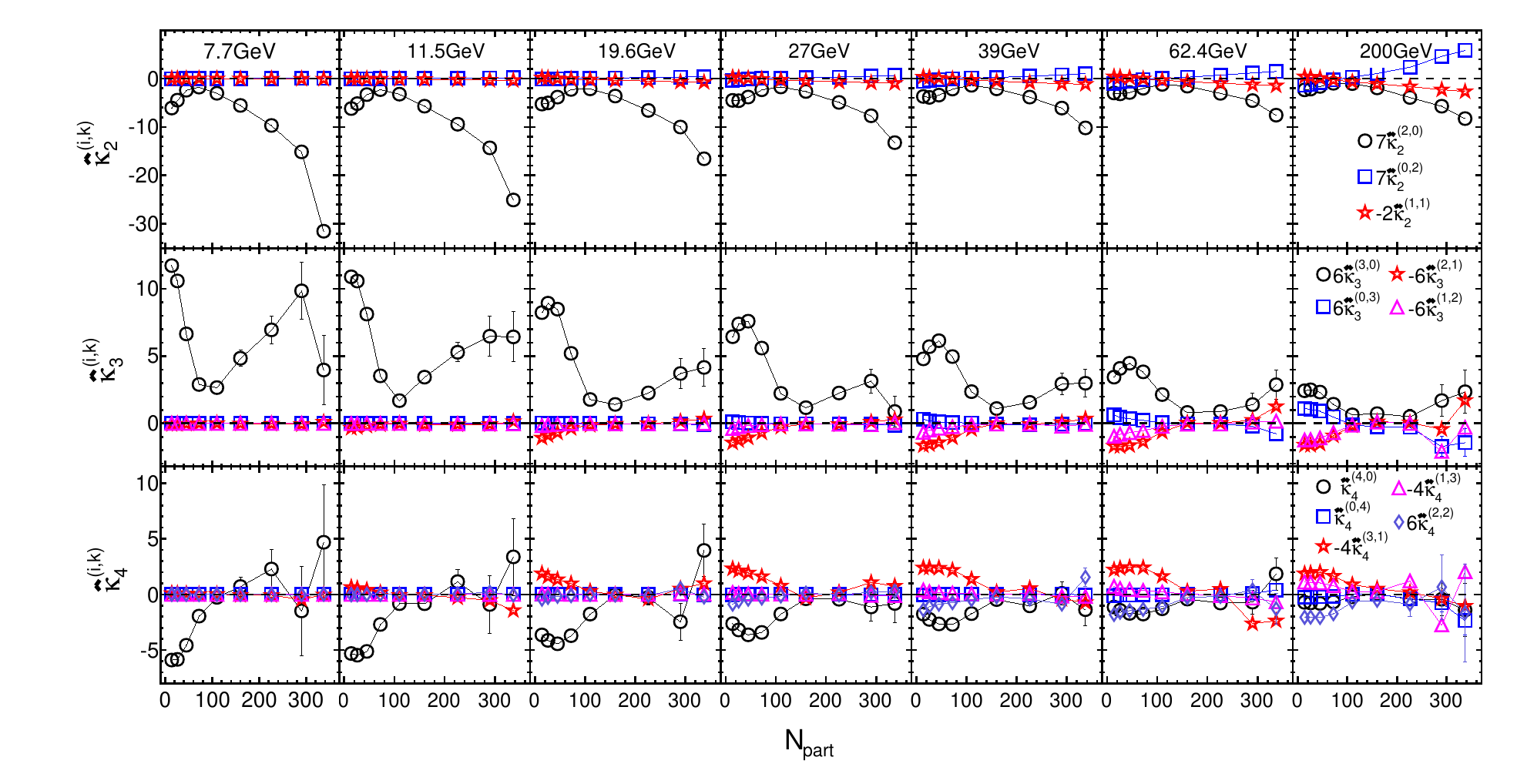}
\caption{ Centrality dependence of the correlation functions of protons (open circle), anti-protons (open square) and various mixed correlations in Au+Au collisions at $\sqrt{s_{NN}}$ = 7.7, 11.5, 19.6, 27, 39, 62.4 and 200 GeV. }
\label{Fig:mix_corr}
\end{figure*}

\begin{figure*}[!htpb]
\hspace{-0.8cm}
\includegraphics[scale=0.75]{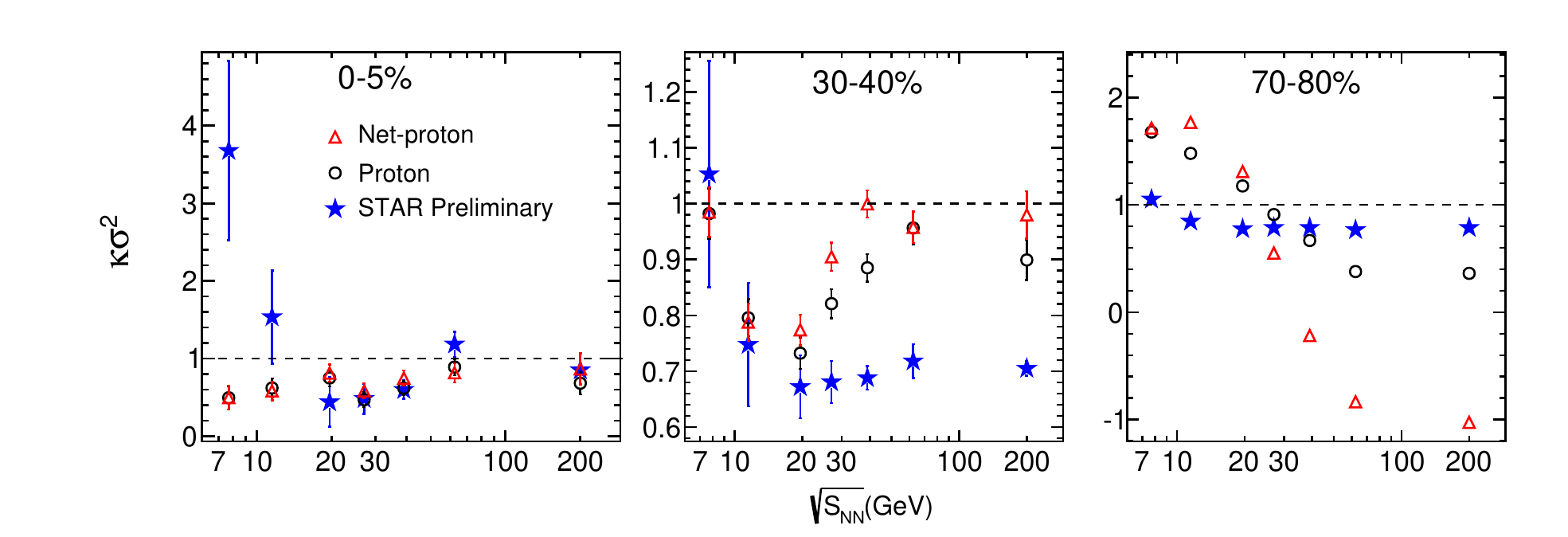}
\caption{Energy dependence of the cumulant ratio ($\kappa\sigma^2$) of net-protons (open triangle) and protons (open circle) in three centrality bins (0-5\%, 30-40\%, 70-80\%) for Au+Au collisions at $\sqrt{s_{NN}}$ =7.7 to 200 GeV from the AMPT model. The solid stars represent the results from STAR experimental data\cite{luo2015energy}. }
\label{Fig:KV}
\end{figure*}

In Fig.\ref{Fig:Net_CF}, we show centrality dependence of various orders of correlation functions for protons and net-protons defined in Eq.\eqref{Eq:correlation} at $\sqrt{s_{NN}}$ =7.7, 11.5, 19.6, 27, 39, 62.4 and 200GeV. From the figure, we observe that the overall trends of correlation functions for protons and net-protons behave similarly with centrality. Their values agree with each other at low RHIC energies, in which one can ignore the contribution of anti-protons~\cite{ab1}.  It can be confirmed from Fig.\ref{Fig:dndy} that the yield of anti-protons is much less than that of protons and thus the $\bar{p}/p$ ratio is vanishingly small in this case. In the first row of the figure, the values of the second-order correlation functions are found to be negative in all centralities and collision energies. The results for net-protons agree well with those for protons when the energies are lower than 200 GeV. At 200 GeV, the values of $(\hat{\kappa}_{2})_{net}$ and $(\hat{\kappa}_{2})_{p}$ separate from each other both at peripheral and at central collisions. For the third-order correlations, we see that $(\hat{\kappa}_{3})_{net}$ and $(\hat{\kappa}_{3})_{p}$ equal  each other in all centralities only at 7.7 and 11.5 GeV. At 19.6 GeV and above, observable differences between them can be found at peripheral collisions. The differences between net-protons and protons become more obvious with increasing energies. In the third row, we observe that $(\hat{\kappa}_{4})_{net}$ are different from $(\hat{\kappa}_{4})_{p}$ from 11.5 GeV and above. From the above behaviors of different order of correlation functions, we infer that the differences between net-protons and protons become larger for higher order correlation functions than for lower ones.

Figure \ref{Fig:p&&Net_CF} shows the second-, third- and fourth-order correlation functions of net-protons as a function of centrality at $\sqrt{s_{NN}}$ = 7.7 - 200 GeV. Note that this plot can reflect the contributions to the fourth-order cumulant from different orders of correlation functions~\eqref{Eq:net_C4} according to the definitions of Eq.~\eqref{Eq:correlation}. In central collisions, when the energy $\sqrt{s_{NN}}\leq 39$ GeV, we find that the magnitudes of $(\hat{\kappa}_{2})_{net}$ are much larger than those of $(\hat{\kappa}_{3})_{net}$ or $(\hat{\kappa}_{4})_{net}$. This means that the second-order correlation function is the dominant contribution to the fourth order net-proton cumulant. When the energy increases up to 62.4 or 200 GeV, various orders of correlation functions play roles comparable to the fourth order cumulant. Therefore, higher order correlations should be considered at higher collision energies. On the other hand, in peripheral collisions, there are no obvious dominant orders of correlation functions at all energies.

In Fig~\ref{Fig:mix_corr}, we compare correlation functions of protons, anti-protons and mixed correlations. Note that we have multiplied correlation functions with the appropriate factors so that they reflect their contribution to the fourth-order cumulant \eqref{Eq:net_C4}. For the second order correlation function, the correlation between protons ($7\hat{\kappa}_{2}^{(2,0)}$) is clearly the dominant part when the energies $\sqrt{s_{NN}}\leq 39$ GeV. The anti-proton correlation ($7\hat{\kappa}_{2}^{(0,2)}$) and mixed correlation ($-2\hat{\kappa}_{2}^{(1,1)}$) are around zero and almost flat in all centralities when $\sqrt{s_{NN}}\leq 39$ GeV. But at 200 GeV, the correlation functions of anti-protons or mixed ones have contribution equivalent to those of protons. This implies that anti-proton number must be considered when calculating correlation functions at high RHIC energies. As the orders of the correlation functions increase up to $3$ or $4$, we observe that the roles of anti-protons and mixed correlations get more important both at central and at peripheral collisions. Thus their contributions should not be neglected in the measurement of higher order correlation functions or cumulants in heavy-ion collisions.

Since there is no published experimental measured correlation function of net-protons available, we can not directly compare our results to the experimental data. In order to make an indirect comparisons with the STAR measured cumulant ratios of net-protons ~\cite{luo2015energy,xiaofenglongpaper}, in Fig.~\ref{Fig:KV} we show the energy dependence of $\kappa\sigma^2$ of protons, net-protons from the AMPT model together with those from the STAR experimental results in three centrality bins. For the AMPT calculations, we first calculated various correlation functions and then obtained cumulants from Eqs.\eqref{Eq:p_C2}-\eqref{Eq:p_C4} for protons and \eqref{Eq:net_C2}-\eqref{Eq:net_C4} for net-protons. In the most central collisions (0-5\%), the results for protons and net-protons from AMPT are almost flat and agree with each other. In the mid-central collisions (30-40\%), they start to separate at energies $\sqrt{s_{NN}}\geq 27$ GeV. In peripheral collisions (70-80\%), the separation gets larger at high energies. This confirms that the contribution of anti-protons is important at high energies and/or in peripheral collisions in the calculations of cumulants or correlation functions. It can be found that, in the most central collisions, the non-monotonic energy dependence of $\kappa\sigma^2$ observed in STAR preliminary results can not be reproduced by the AMPT model. This is because there is no critical physics implemented in the transport model. We also observe that the model can not describe the behaviors of the STAR data either in the mid-central or in the peripheral collisions.

%%%%%%%%%%%%%%%%%%%%%%%%%%%%%%%%%%%%%%%%%%%%%%%%%%%%%%%%%%%%%%%%%%%%%%%%%%%%%%%
\section{Summary}
In this paper we have studied various orders of correlation functions of net-proton in Au+Au collisions at $\sqrt{s_\mathrm{NN}}$=7.7 to 200 GeV. It is the first time that the contribution of anti-protons has been taken into account together with that of proton in calculating correlation functions by using the AMPT model.

We find that the correlation functions from net-protons and only protons agree with each other in low RHIC energies. But at high energies, they have a clear separation, especially for peripheral collisions. We infer that the contribution of anti-protons is important at high RHIC energies and/or in peripheral collisions. In central collisions, the magnitude of net-proton correlation functions of the second-order is much larger than those of the higher orders, while they play comparable roles with increasing energies up to 200 GeV. For the centrality dependence of mixed correlations, we observe that proton correlation is the dominant part at low energies, while at high energies, the correlations between anti-protons and the mixed correlations have equivalent contributions to those of protons. It is found that, in the calculation of correlation functions, the contribution from anti-protons is more important for higher order correlations than for lower ones. The comparison of the cumulants between experimental data and the AMPT calculations shows that the STAR preliminary results can not be described by the AMPT model without implementing critical physics.

The STAR experiment has planned a second phase of the beam energy scan (BES-II) program~\cite{besII,besII2} to run in 2019 - 2020. One of the primary goals of BES-II is the search for evidence of a phase transition between hadronic gas and QGP phases. With significant improved statistics and particle identification in BES-II, it would be of great interests if STAR could measure the correlation functions of net-protons together with high-order cumulant ratios to explore the QCD phase diagram. The results from  AMPT model calculations here can provide non-critical estimations for the background contributions to the QCD critical point search in heavy-ion collisions.

\section*{Acknowledgments}
We thank Prof. Yuangfang Wu and Xiaofeng Luo for useful discussions and comments. We further thank the STAR Collaboration for providing us with their preliminary data. This work is supported in part by the Ministry of Science and Technology (MoST) under Grant No. 2016YFE0104800, the Major State Basic Research Development Program of China under Grant No. 2014CB845402, the NSFC of China under Grant No. 11405088, and the Chinese Scholarship Council No. 201506775038.

%\bibliography{ref}% BiTex form

\begin{thebibliography}{99}
\bibitem{StephanovPD} M. A. Stephanov, K. Rajagopal, and E. V. Shuryak, Phys. Rev. Lett. 81, 4816 (1998).
\bibitem{adams2005experimental} J. Adams  {\it et al.} (STAR Collaboration), Nucl. Phys. A757, 102 (2005).
\bibitem{conservecharge0}M. Asakawa, U. W. Heinz, and B. Muller, Phys. Rev. Lett. 85, 2072 (2000).
\bibitem{conservecharge1} V. Koch, A. Majumder, and J. Randrup, Phys. Rev. Lett. 95, 182301 (2005).
\bibitem{ejiri2006hadronic} S. Ejiri, F. Karsch, and K. Redlich, Phys. Lett. B 633, 275 (2006).
\bibitem{conservecharge2} M. A. Stephanov, Phys. Rev. Lett. 102, 032301 (2009).
\bibitem{conservecharge3} M. A. Stephanov,  Phys. Rev. Lett. 107, 052301 (2011).
\bibitem{conservecharge4} B. J. Schaefer and M. Wagner, Phys. Rev. D85, 034027 (2012).
\bibitem{asakawa2009third} M. Asakawa, S. Ejiri, and M. Kitazawa, Phys. Rev. Lett. 103, 262301 (2009).
\bibitem{net_proton2010} M. M. Aggarwal {\it et al.} (STAR Collaboration), Phys. Rev. Lett. 105, 022302 (2010).
\bibitem{net_proton2014} L. Adamczyk {\it et al.} (STAR Collaboration), Phys. Rev. Lett. 112, 032302 (2014).
\bibitem{net_charge2014} L. Adamczyk {\it et al.} (STAR Collaboration), Phys. Rev. Lett. 113, 092301 (2014).
\bibitem{xu2017_net-kaon} J. Xu (for the STAR Collaboration), J. Phys. G: Conf. Seri. 736, 012002 (2016).
\bibitem{amal2017_net-kaon} A. Sarkar (for the STAR Collaboration), J. Phys. G: Conf. Seri. 509, 012069 (2014).
\bibitem{STARnet_kaon2017} L. Adamczyk {\it et al.} (STAR Collaboration), arXiv:1709.00773.
\bibitem{Jochen_star} J. Thader (for the STAR Collaboration), Nucl. Phys. A 956, 320 (2016).
\bibitem{phenix} A. Adare  {\it et al.} (PHENIX Collaboration), Phys. Rev. C 93, 011901(R) (2016).
\bibitem{luo2015energy} X. Luo (for the STAR Collaboration), in Proceedings of the 9th International Workshop on Critical Point and Onset of Deconfinement
(CPOD 2014), Bielefeld, Germany, November 17\_21, 2014, [PoS CPOD2014, 019 (2015)].
\bibitem{xiaofenglongpaper} X. Luo and N. Xu, Nucl. Sci. Technol. 28, 112 (2017), arXiv:1701.02105.
\bibitem{BLing} B. Ling and M. A. Stephanov, Phys. Rev. C 93, 034915 (2016).
\bibitem{ab1} A. Bzdak, V. Koch, and N. Strodthoff, Phys. Rev. C 95, 054906 (2017).
\bibitem{ab2} A. Bzdak,V.Koch,V. Skokov, andN. Strodthoff, in Proceedings of the XXVI International Conference on Ultrarelativistic Heavy-Ion Collisions (Quark Matter 2017), Chicago, USA, February 5¨C11, 2017 [Nucl. Phys. A 967, 465 (2017)].
\bibitem{ab22} A. Bzdak and V. Koch, arXiv:1707.02640.
\bibitem{ab3} A. Bzdak and V. Koch,  Phys. Rev. C 86, 044904 (2012).

\bibitem{Roli} R. Esha (for the STAR Collaboration), in Proceedings of the XXVI International Conference on Ultrarelativistic Heavy-Ion Collisions (QuarkMatter 2017), Chicago, USA, February 5¨C11, 2017 [Nucl. Phys. A 967, 457 (2017)].
\bibitem{luorqmd} S. He and X. Luo,  arXiv:1704.00423.
\bibitem{bzdak2016multiparticle}A. Bzdak and P. Bo{\.z}ek,  Phys. Rev. C 93, 024903 (2016).
\bibitem{asymmetry1}  A. Bzdak and K. Dusling, Phys. Rev. C 93, 031901(R) (2016).
\bibitem{asymmetry2}  A. Bzdak and K. Dusling, Phys. Rev. C 94, 044918 (2016).
\bibitem{pbarpratio} L. Adamczyk {\it et al.} (STAR Collaboration), arXiv:1701.07065.
\bibitem{mix} A. Bzdak, V. Koch, and V. Skokov, Eur. Phys. J. C 77£¬ 288 (2017).
\bibitem{relation1} M. Kitazawa and X. Luo, Phys. Rev. C 96, 024910 (2017).
\bibitem{relation2} T. Nonaka, M. Kitazawa and S. Esumi, Phys. Rev. C 95, 064912 (2017).
\bibitem{lin2005AMPT} Z. W. Lin, C. M. Ko, B. A. Li, B. Zhang, and S. Pal, Phys. Rev. C 72, 064901 (2005).
\bibitem{lin2014recent} Z. W. Lin, arXiv:1403.1854.
\bibitem{CBWC} X. Luo, J. Phys. G: Nucl. Part. Phys. 39, 025008 (2012).
\bibitem{efron1993introductionBootstrap} B. Efron and R. J. Tibshirani, An Introduction to the Bootstrap (CRC, Boca Raton, FL, 1994), p. 436.
\bibitem{efron1986introductionBootstrap} B. Efron and R. Tibshirani, Stat. Sci. 1, 54 (1986).
\bibitem{besII} STAR Collaboration, STAR Note 598, https://drupal.star.bnl.gov/STAR/starnotes/public/sn0598.
\bibitem{besII2} C. Yang (for the STAR Collaboration), in Proceedings of the XXVI International Conference on Ultrarelativistic Heavy-Ion Collisions (QuarkMatter 2017), Chicago, USA, February 5¨C11, 2017 [Nucl. Phys. A 967, 800 (2017)].
\end{thebibliography}
%\include{ref}
%%%%%%%%%%%%%%%%%% Citation %%%%%%%%%%%%%%%%%%%%%%%%%

\end{document}